\documentclass[11pt]{article}
\usepackage{amssymb,amsmath,amsfonts}
\usepackage{graphicx}
\usepackage{graphics}
\usepackage{eepic,epsfig}

\begin{document}

\title{Vacuum densities in braneworlds\thanks{%
Talk presented at the First International Congress of Armenian
Physicists, 11-15 September, 2005, Yerevan, Armenia}}
\author{Aram A. Saharian\thanks{%
Email: saharyan@server.physdep.r.am} \\
\textit{Department of Physics, Yerevan State University, 1 Alex Manoogian
Str.} \\
\textit{\ 375025 Yerevan, Armenia}}
\date{\today}
\maketitle

\begin{abstract}
We give a short review of the recent development in the
investigation of the vacuum expectation values of the bulk and
surface energy-momentum tensors generated by quantum fluctuations
of a massive scalar field with general curvature coupling
parameter subject to Robin boundary conditions on two codimension
one parallel branes located on $(D+1)$-dimensional anti-de Sitter
(AdS) bulk. An application to the Randall-Sundrum braneworld with
arbitrary mass terms on the branes is discussed.
\end{abstract}

\bigskip


\bigskip

\section{Introduction}

Recent proposals of large extra dimensions use the concept of
brane as a sub-manifold embedded in a higher dimensional
spacetime, on which the Standard Model particles are confined.
Braneworlds naturally appear in string/M-theory context and
provide a novel setting for discussing phenomenological and
cosmological issues related to extra dimensions (for reviews in
braneworld gravity and cosmology see Refs. \cite{Ruba01}). The
investigation of quantum effects in braneworld models is of
considerable phenomenological interest, both in particle physics
and in cosmology. The braneworld corresponds to a manifold with
boundaries and all fields which propagate in the bulk will give
Casimir-type contributions to the vacuum energy, and as a result
to the vacuum forces acting on the branes. In dependence of the
type of a field and boundary conditions imposed, these forces can
either stabilize or destabilize the braneworld. In addition, the
Casimir energy gives a contribution to both the brane and bulk
cosmological constants and, hence, has to be taken into account in
the self-consistent formulation of the braneworld dynamics.
Motivated by these, the role of quantum effects in braneworld
scenarios has received a great deal of attention (see, for
instance, references given in \cite{Saha04a}). In particular, the
investigation of local physical characteristics, such as
expectation values of the bilinear field products and the
energy-momentum tensor, is of considerable interest. In addition
to describing the physical structure of the quantum field at a
given point, the energy-momentum tensor acts as the source in the
Einstein equations and therefore plays an important role in
modelling a self-consistent dynamics involving the gravitational
field. In this paper based on Refs. \cite{Saha04a,Saha04b}, we
give a short review of the recent progress in investigations of
the vacuum expectation values (VEVs) for both bulk and surface
energy-momentum tensors in braneworlds with AdS bulk (for the
vacuum energy-momentum tensor in the Randall-Sundrum braneworld
see also \cite{Knap03}). The corresponding results for the
geometry of two codimension one parallel branes located on
$(D+1)$-dimensional background spacetime $AdS_{D_1+1}\times \Sigma
$ with a warped internal space $\Sigma $ are presented in
\cite{Saha05a}.

\section{Bulk energy-momentum tensor and vacuum interaction forces}

Consider a massive scalar field $\varphi (x)$ on background of a $(D+1)$%
-dimensional AdS spacetime ($AdS_{D+1}$) with the line element
\begin{equation}
ds^{2}=g_{ik}dx^{i}dx^{k}=e^{-2k_{D}y}\eta _{\mu \nu }dx^{\mu }dx^{\nu
}-dy^{2},  \label{metric}
\end{equation}%
and AdS radius given by $1/k_{D}$. Here $\eta _{\mu \nu }=\mathrm{diag}%
(1,-1,\ldots ,-1)$ is the metric for the $D$-dimensional Minkowski
spacetime, $i,k=0,1,\ldots ,D$, and $\mu ,\nu =0,1,\ldots ,D-1$. The
corresponding field equation reads
\begin{equation}
\left([ \nabla _{i}\nabla ^{i}+m^{2}-\zeta D(D+1)k_{D}^{2}\right]
\varphi (x)=0,  \label{fieldeq}
\end{equation}%
where the symbol $\nabla _{i}$ is the operator for the covariant
derivative, $\zeta $ is the curvature coupling parameter. We will
assume that the field obeys Robin boundary conditions on two
parallel infinite plane boundaries (branes), located at $y=a$ and
$y=b$, $a<b$:
\begin{equation}
\left( \tilde{A}_{y}+\tilde{B}_{y}\partial _{y}\right) \varphi (x)=0,\quad
y=a,b,  \label{boundcond}
\end{equation}%
with constant coefficients $\tilde{A}_{y}$, $\tilde{B}_{y}$. The
eigenfunctions of the problem are decomposed into standard plane
waves with the wave vector $\mathbf{k}$ in directions parallel to
the branes and into the
transverse part which is a cylindrical function of the argument $uz$, where $%
z=e^{k_{D}y}/k_{D}$. The eigenvalues $u=u_{\nu ,n}$, $n=1,2,\ldots $, are
solutions to the equation
\begin{equation}
g_{\nu }^{(ab)}(uz_{a},uz_{b})\equiv \bar{J}_{\nu }^{(a)}(uz_{a})\bar{Y}%
_{\nu }^{(b)}(uz_{b})-\bar{Y}_{\nu }^{(a)}(uz_{a})\bar{J}_{\nu
}^{(b)}(uz_{b})=0,  \label{cnu}
\end{equation}%
where $J_{\nu }(x)$, $Y_{\nu }(x)$ are the Bessel and Neumann functions of
the order
\begin{equation}
\nu =\sqrt{(D/2)^{2}-D(D+1)\zeta +m^{2}/k_{D}^{2}}.  \label{nu}
\end{equation}%
In (\ref{cnu}), $z_{j}=e^{k_{D}j}/k_{D}$, $j=a,b$, and the
notation
\begin{equation}
\bar{F}^{(j)}(x)=A_{j}F(x)+B_{j}xF^{\prime }(x),  \label{notbar}
\end{equation}%
is introduced for a given function $F(x)$, with the coefficients
\begin{equation}
A_{j}=\tilde{A}_{j}+\tilde{B}_{j}k_{D}D/2,\quad B_{j}=\tilde{B}_{j}k_{D}.
\label{AAtilda}
\end{equation}%
The VEVs of the physical quantities bilinear in the field operator
(Wightman function, field square, energy-momentum tensor) are
expressed in terms of the mode-sums which contain the summation
over the eigenvalues $u_{\nu ,n}$. The application of the
generalized Abel-Plana formula \cite{SahaAP} to this sum allows us
to extract the boundary-free AdS part and to present the boundary
induced parts in terms of exponentially convergent integrals. In
particular, the VEV of the energy-momentum tensor in the region
between the branes is presented in the form
\begin{equation}
\langle 0|T_{i}^{k}|0\rangle =\langle 0_{S}|T_{i}^{k}|0_{S}\rangle +\langle
T_{i}^{k}\rangle ^{(j)}-\frac{k_{D}^{D+1}z^{D}\delta _{i}^{k}}{2^{D-1}\pi ^{%
\frac{D}{2}}\Gamma \left( \frac{D}{2}\right) }\int_{0}^{\infty
}duu^{D-1}\Omega _{j\nu }(uz_{a},uz_{b})F^{(i)}\left[ G_{\nu
}^{(j)}(uz_{a},uz)\right] ,  \label{Tik1int}
\end{equation}%
where $j=a,b$ provide two alternative representations. In this formula $%
\langle 0_{S}|T_{i}^{k}|0_{S}\rangle $ is the VEV\ for the AdS bulk without
boundaries and the part $\langle T_{i}^{k}\rangle ^{(j)}$ is induced by a
single brane located at $y=j$ when the second brane is absent. For $j=a$ the
single brane induced part is given by formula
\begin{equation}
\langle T_{i}^{k}\rangle ^{(a)}=-\frac{k_{D}^{D+1}z^{D}\delta _{i}^{k}}{%
2^{D-1}\pi ^{\frac{D}{2}}\Gamma \left(
\frac{D}{2}\right)}\int_{0}^{\infty }duu^{D-1}\frac{\bar{I}_{\nu
}^{(a)}(uz_{a})}{\bar{K}_{\nu }^{(a)}(uz_{a})}F^{(i)}\left[ K_{\nu }(uz)%
\right] ,\quad z>z_{a},  \label{Tik1plnew}
\end{equation}%
where $I_{\nu }(z)$ and $K_{\nu }(z)$ are the Bessel modified
functions.
The corresponding formula for $j=b$ and $z<z_{b}$ is obtained from (\ref%
{Tik1plnew}) by the replacements $I\rightleftarrows K$. The last term on the
right of formula (\ref{Tik1int}) is induced by the second brane. In the
corresponding formula we use the notations
\begin{eqnarray}
F^{(i)}[g(v)] &=&\left( \frac{1}{2}-2\zeta \right) \left[ v^{2}g^{\prime
2}(v)+\left( D+\frac{4\zeta }{4\zeta -1}\right) vg(v)g^{\prime }(v)+\right.
\notag \\
&&+\left. \left( \nu ^{2}+v^{2}+\frac{2v^{2}}{D(4\zeta -1)}\right) g^{2}(v)%
\right] ,\quad i=0,1,\ldots ,D-1,  \label{Finew} \\
F^{(D)}[g(v)] &=&-\frac{v^{2}}{2}g^{\prime }{}^{2}(v)+\frac{D}{2}\left(
4\zeta -1\right) vg(v)g^{\prime }(v)+  \nonumber \\
&&+\frac{1}{2}\left[ v^{2}+\nu ^{2}+2\zeta D(D+1)-D^{2}/2\right]
g^{2}(v),\label{FDnew}
\end{eqnarray}%
for a given function $g(v)$,
\begin{eqnarray}
G_{\nu }^{(j)}(u,v) &=&I_{\nu }(v)\bar{K}_{\nu }^{(j)}(u)-K_{\nu }(v)\bar{I}%
_{\nu }^{(j)}(u),  \label{Gnu} \\
\Omega _{a\nu }(u,v) &=&\frac{\bar{K}_{\nu }^{(b)}(v)/\bar{K}_{\nu }^{(a)}(u)%
}{\bar{K}_{\nu }^{(a)}(u)\bar{I}_{\nu }^{(b)}(v)-\bar{K}_{\nu }^{(b)}(v)\bar{%
I}_{\nu }^{(a)}(u)}, \label{Oma}
\end{eqnarray}
and the corresponding expression for $\Omega _{b\nu }(u,v)$ is obtained from formula (%
\ref{Oma}) by the replacement $\bar{K}_{\nu
}^{(b)}(v)/\bar{K}_{\nu }^{(a)}(u)\rightarrow \bar{I}_{\nu
}^{(a)}(u)/\bar{I}_{\nu }^{(b)}(v)$. As we could expect from the
problem symmetry the brane induced parts of the vacuum
energy-momentum tensor corresponding to the components on the
hyperplane parallel to the branes are proportional to the
corresponding metric tensor. On the boundary the vacuum
energy-momentum tensor for a single brane diverges, except the
case of a conformally coupled massless scalar. The leading terms
of the corresponding asymptotic expansion near the boundary are
the same as for a plate in the Minkowski bulk. For large proper
distances from the brane compared with the AdS curvature radius, $%
k_{D}|y-a|\gg 1$, the single brane induced energy-momentum tensor vanishes
as $\exp [2\nu k_{D}(a-y)]$ in the region $y>a$ and as $\exp [k_{D}(2\nu
+D)(y-a)]$ in the region $y<a$. The same behavior takes place for a fixed $%
y-a$ and large values of the parameter $k_{D}$. In the large mass limit, $%
m\gg k_{D}$, the boundary parts are exponentially suppressed. For
a conformally coupled massless scalar field one has $\nu =1/2$ and
the corresponding cylindrical functions are expressed through the
elementary functions. In this case the corresponding VEVs can also
be obtained from the results for two parallel plates in the
Minkowski bulk by using the standard conformal transformation
technique \cite{Saha03a}.

For the case of two-branes geometry the vacuum force acting per unit surface
of the brane at $y=j$ is determined by the ${}_{D}^{D}$ -- component of the
vacuum energy-momentum tensor at this point. The corresponding effective
pressures can be presented as a sum of two terms: $p_{j}=p_{j1}+p_{j{\mathrm{%
(int)}}}$, $j=a,b$, where $p_{j1}$ is the pressure for a single
brane at $y=j $ when the second brane is absent. This term is
divergent due to the surface divergences in the vacuum expectation
values and needs an additional renormalization. The term
\begin{eqnarray}
p_{j{\mathrm{(int)}}} &=&\frac{k_{D}^{D+1}}{2^{D}\pi
^{\frac{D}{2}}\Gamma \left( \frac{D}{2}\right) }\int_{0}^{\infty
}dx\,x^{D-1}\Omega _{j\nu }\left(
xz_{a}/z_{j},xz_{b}/z_{j}\right)   \notag \\
&&\times \left\{ \left( x^{2}-\nu ^{2}+2m^{2}/k_{D}^{2}\right)
B_{j}^{2}-D(4\zeta -1)A_{j}B_{j}-A_{j}^{2}\right\} ,  \label{pintj}
\end{eqnarray}%
is the pressure induced by the presence of the second brane, and
can be termed as an interaction force. For Dirichlet scalar these
forces are always attractive. In the case of Robin boundary
conditions the interaction forces can be either repulsive or
attractive. Moreover, there is a region in the space of Robin
parameters in which the interaction forces are repulsive for small
distances and are attractive for large distances. This provides a
possibility to stabilize interbrane distance by using the vacuum
forces. For large distances between the branes, the vacuum
interaction forces per unit surface are exponentially suppressed
by the factor $\exp [2\nu k_{D}(a-b)]$ for the brane at $y=a$ and
by the factor $\exp [k_{D}(2\nu +D)(a-b)]$ for the brane at $y=b$.
In the Randall-Sundrum braneworld with arbitrary mass terms
$c_{j}$, $j=a,b$, on the branes the Robin coefficients are
expressed
through these mass terms and the curvature coupling parameter as%
\begin{equation}
\tilde{A}_{j}/\tilde{B}_{j}=-\frac{1}{2}(n_{j}c_{j}+4D\zeta
k_{D}),\;n_{a}=1,\;n_{b}=-1.  \label{ABtilde}
\end{equation}
For the twisted scalar Dirichlet boundary conditions are obtained
on both branes. The corresponding VEV of the energy-momentum
tensor is defined by formulae (\ref{Tik1int}), (\ref{Tik1plnew})
by an additional factor 1/2.

\section{Surface densities}

In the previous section we have considered the bulk
energy-momentum tensor. For a scalar field on manifolds with
boundaries in addition to this part, the energy-momentum tensor
contains a contribution located on the boundary. The surface part
of the energy-momentum tensor is essential in considerations of
the relation between local and global characteristics in the
Casimir effect. For an arbitrary smooth boundary $\partial M_{s}$
with the inward-pointing unit normal vector $n^{l}$, the surface
part of the
energy-momentum tensor is given by the formula \cite{Saha03} $T_{ik}^{%
\mathrm{(surf)}}=\delta (x;\partial M_{s})\tau _{ik}$ with
\begin{equation}
\tau _{ik}=\zeta \varphi ^{2}K_{ik}-(2\zeta -1/2)h_{ik}\varphi n^{l}\nabla
_{l}\varphi ,  \label{tausurf}
\end{equation}%
where $K_{ik}$ is the extrinsic curvature tensor of the boundary
and $h_{ik}$ is the corresponding induced metric. In the problem
under consideration the VEV of the surface energy-momentum tensor
on the brane $y=j$ is presented in the form
\begin{equation}
\langle 0|\tau _{\mu \nu }^{(j)}|0\rangle =-g_{\mu \nu }n_{j}\left[ \zeta
k_{D}-(2\zeta -1/2)\tilde{A}_{j}/\tilde{B}_{j}\right] \langle 0|\varphi
^{2}|0\rangle _{z=z_{j}},  \label{tauj}
\end{equation}%
and $\langle 0|\tau _{DD}^{(j)}|0\rangle =0$. From the point of view of
physics on the brane, Eq. (\ref{tauj}) corresponds to the gravitational
source of the cosmological constant type with the equation of state $%
\varepsilon _{j}^{{\mathrm{(surf)}}}=-p_{j}^{{\mathrm{(surf)}}}$.
By making use of expressions (\ref{ABtilde}), it can be seen that
the surface energy in the Randall-Sundrum braneworld vanishes for
minimally and conformally coupled scalar fields with zero brane
mass terms. The VEV (\ref{tauj}) diverges and needs some
regularization. We use the method which is an analog of the
generalized zeta function approach. The VEV\ of the field square
is presented in terms of the generalized zeta function
\begin{equation}
\zeta _{j}(s)=\sum_{n=1}^{\infty }\frac{u_{\nu ,n}^{D+s}g_{\nu
}^{(l)}(u_{\nu ,n}z_{l},u_{\nu ,n}z_{j})}{\frac{\partial }{\partial u}g_{\nu
}^{(ab)}(uz_{a},uz_{b})|_{u=u_{\nu ,n}}},  \label{zetsx}
\end{equation}
with $j,l=a,b$, $l\neq j$, $g_{\nu }^{(l)}(u,v)=J_{\nu
}(v)\bar{Y}_{\nu }^{(l)}(u)-Y_{\nu }(v)\bar{J}_{\nu }^{(l)}(u)$,
as
\begin{equation}
\langle 0|\varphi ^{2}|0\rangle _{z=z_{j}}=k_{D}^{D-1}z_{j}^{D}\frac{%
B_{j}\beta _{D}}{\mu ^{1+s}}B\left( \frac{D-1}{2},-\frac{D-1+s}{2}\right)
\zeta _{j}(s)|_{s=-1},  \label{IFs0}
\end{equation}%
where  $|_{s=-1}$ is understood in the sense of the analytic continuation,
and $B(x,y)$ is the Euler beta function. Using the Cauchy's theorem on
residues, the following integral representation for the zeta functions is
obtained
\begin{equation}
\zeta _{j}(s)=-\frac{1}{\pi }\sin \frac{\pi }{2}\left( D+1+s\right)
\int_{0}^{\infty }du\,u^{D+s}\left[ n_{j}U_{j\nu }(uz_{j})+B_{j}\Omega
_{j\nu }(uz_{a},uz_{b})\right] ,  \label{zet12}
\end{equation}
with $U_{a\nu }(x)=K_{\nu }(x)/\bar{K}_{\nu }^{(a)}(x)$ and $U_{b\nu
}(x)=I_{\nu }(x)/\bar{I}_{\nu }^{(b)}(x)$. The contribution of the second
term in the square brackets is finite at $s=-1$ and vanishes in the limits $%
z_{a}\rightarrow 0$ and $z_{b}\rightarrow \infty $. The first term in the
square brackets of this expression corresponds to the contribution of a
single brane at $z=z_{j}$ when the second brane is absent. The
regularization is needed for this term only and the details for the
corresponding analytic continuation can be found in \cite{Saha04b}. For the
analytic continuation of the single brane zeta functions we subtract and add
to the integrands leading terms of the corresponding asymptotic expansions,
and present them as sums of two parts. The first one is convergent at the
physical point and can be evaluated numerically. In the second, asymptotic
part the pole contributions are given explicitly. As a consequence, the
single brane surface Casimir energies for separate left and right sides of
the brane contain pole and finite contributions. The remained pole term is a
characteristic feature for the zeta function regularization method and has
been found in the calculations of the total Casimir energy for many cases of
boundary geometries. As in the case of the total vacuum energy, the
renormalization of these terms can be performed by using the brane tensions.
For an infinitely thin brane taking the left and right regions together, in
odd spatial dimensions the pole parts cancel and the surface Casimir energy
is finite. In this case the total surface energy per unit physical volume on
the brane (surface tension) does not depend on the brane position. The
results of the corresponding numerical evaluation show that, in dependence
of ratio of coefficients in the boundary condition, the surface energy for a
single brane can be either negative or positive. In two branes geometry the
surface energy density on the brane at $z=z_{j}$ is presented as the sum
\begin{equation}
\varepsilon _{j}^{{\mathrm{(surf)}}}=\varepsilon _{j1}^{{\mathrm{(surf)}}%
}+\Delta \varepsilon _{j}^{{\mathrm{(surf)}}},  \label{emt2pl2}
\end{equation}%
where $\varepsilon _{j1}^{{\mathrm{(surf)}}}$ is the surface energy density
induced on the corresponding surface of a single brane at $y=j$ when the
second brane is absent, and the term
\begin{equation}
\Delta \varepsilon _{j}^{{\mathrm{(surf)}}}=\left[ 2\zeta B_{j}-\left(
4\zeta -1\right) \tilde{A}_{j}\right] n^{(j)}(k_{D}z_{j})^{D}B_{j}\beta
_{D+1}\int_{0}^{\infty }du\,u^{D-1}\Omega _{j\nu }(uz_{a},uz_{b})
\label{emt2pl3}
\end{equation}%
is the energy density induced by the presence of the second brane.
Note that this part is located on the surface $y=a+0$ for the
brane at $y=a$ and on
the surface $y=b-0$ for the brane at $y=b$. On the surfaces $y=a-0$ and $%
y=b+0$ the surface densities are the same as for single branes. The
expression on the right of Eq. (\ref{emt2pl3}) is finite for all values $%
z_{a}<z_{b}$ and is a function on the interbrane distance only. In the limit
of large AdS radius, $k_{D}\rightarrow 0$, the result for two parallel
plates on the Minkowski bulk is recovered. For large distances between the
branes, the surface densities induced by the second brane are exponentially
suppressed by the factor $\exp [k_{D}(2\nu +D)(a-b)]$ for the brane at $y=a$
and by the factor $\exp [2\nu k_{D}(a-b)]$ for the brane at $y=b$. The
exponential suppression also takes place in the large mass limit. From the
viewpoint of an observer living on the brane at $y=j$, the $D$-dimensional
Newton's constant is related to the higher-dimensional fundamental Newton's
constant by formula
\begin{equation}
G_{Dj}=\frac{(D-2)k_{D}G_{D+1}}{e^{(D-2)k_{D}(b-a)}-1}e^{(D-2)k_{D}(b-j)}.
\label{GDj}
\end{equation}%
In the orbifolded version of the model an additional factor 2 appears in the
denominator of the expression on the right. For large interbrane separations
this constant is exponentially small on the brane $y=b$ and the
gravitational interactions on this brane are exponentially suppressed. The
corresponding effective cosmological constant generated by the second brane
is determined by the relation
\begin{equation}
\Lambda _{Dj}=8\pi G_{Dj}\Delta \varepsilon _{j}^{{\mathrm{(surf)}}}=\frac{%
8\pi \Delta \varepsilon _{j}^{{\mathrm{(surf)}}}}{M_{Dj}^{D-2}},
\label{effCC}
\end{equation}
where $M_{Dj}$ is the $D$-dimensional effective Planck mass scale for an
observer on the brane. For large interbrane distances the quantity (\ref%
{effCC}) is suppressed to compared with the corresponding Planck scale
quantity in the brane universe by the factor $\exp [k_{D}(2\nu +D)(a-b)]$,
assuming that the AdS inverse radius and the fundamental Planck mass are of
the same order. In the original Randall-Sundrum model with $D=4$, for a
scalar field with the mass $|m^{2}|\lesssim k_{D}^{2}$, and interbrane
distances solving the hierarchy problem, the value of the cosmological
constant on the visible brane by order of magnitude is in agreement with the
value suggested by current cosmological observations without an additional
fine tuning of the parameters.


\begin{thebibliography}{99}

\bibitem{Ruba01} V.A. Rubakov, Phys. Usp. {\bf 44}, 871 (2001); R. Maartens, gr-qc/0312059.

\bibitem{Saha04a} A.A. Saharian, Nucl. Phys. \textbf{B712}, 196 (2005).

\bibitem{Saha04b} A.A. Saharian, Phys. Rev. D \textbf{70}, 064026 (2004).

\bibitem{Knap03} A. Knapman and D.J. Toms, Phys. Rev. D \textbf{69}, 044023
(2004).

\bibitem{Saha05a} A.A. Saharian, hep-th/0508038; hep-th/0508185.

\bibitem{SahaAP} A.A. Saharian, "The generalized Abel-Plana formula. Applications
to Bessel functions and Casimir effect," Report No. IC/2000/14, hep-th/0002239.

\bibitem{Saha03a} A.A. Saharian and M.R. Setare, Phys. Lett. B \textbf{552},
119 (2003).

\bibitem{Saha03} A.A. Saharian, Phys. Rev. D \textbf{69}, 085005 (2004).

\end{thebibliography}
\end{document}